\documentclass[aps,prl,twocolumn,10pt,superscriptaddress,nofootinbib,longbibliography,floatfix]{revtex4-2}

\usepackage[T1]{fontenc}
\usepackage[utf8]{inputenc}
\usepackage{amsmath,amssymb,bm}
\usepackage{graphicx}
\usepackage{xcolor}
\usepackage{placeins}
\usepackage{hyperref}

\hypersetup{colorlinks=true,linkcolor=blue,citecolor=blue,urlcolor=blue}

\newcommand{\deltalambda}{\delta_\lambda}
\newcommand{\hini}{h_{\mathrm{ini}}}

\begin{document}

\title{Critical Touching of Temporal Entanglement Transitions}

\author{Ting-Long Wang} 
\affiliation{State Key Laboratory of Quantum Functional Materials, School of Physical Science and Technology, ShanghaiTech University, Shanghai 201210, China} 
\author{Shi-Xin Zhang} 
\affiliation{Institute of Physics, Chinese Academy of Sciences, Beijing 100190, China} 
\author{Shuai Yin}
\email{yinsh6@mail.sysu.edu.cn} 
\affiliation{Guangdong Provincial Key Laboratory of Magnetoelectric Physics and Devices, School of Physics, Sun Yat-sen University, Guangzhou 510275, China} 
\affiliation{School of Physics, Sun Yat-sen University, Guangzhou 510275, China} 
\author{Yi-Fan Jiang} 
\email{jiangyf2@shanghaitech.edu.cn}
\affiliation{State Key Laboratory of Quantum Functional Materials, School of Physical Science and Technology, ShanghaiTech University, Shanghai 201210, China}

\date{September 20, 2026}

\begin{abstract}
Equilibrium phase transitions are conventionally categorized into first‑order and continuous phase transitions. Far from equilibrium, many new transitions emerge during the real-time evolution of quantum systems. One such transition is the temporal entanglement transition (TET) characterized by the nonanalyticity of the entanglement spectrum. So far, all TETs occur through a linear crossing of the leading Schmidt levels in different symmetry sectors, resembling a first-order transition in equilibrium. 
A natural question is whether a continuous TET, featured by entanglement spectrum touching, is possible. In a periodically driven transverse $J_1$-$J_2$ Ising chain, we show that two such TETs can merge into a critical touching, where the leading levels meet tangentially without exchanging, realizing the temporal analog of a continuous phase transition. 
Near the critical frequency, the temporal separation of the two TETs vanishes continuously and its derivative with respect to frequency diverges, a nonanalytic signature that is absent in a single first-order TET. This finite-frequency touching arises from the interplay between a weak symmetry-preserving perturbation of the product initial state and Floquet corrections. Further extending the frequency scan reveals a second critical touching at a higher frequency, and the two critical frequencies enclose a finite window with no TET. These features can be understood from a second-order Floquet Hamiltonian and persist across a broad range of coupling ratios, establishing the critical touching as a distinct form of TET.
\end{abstract}

\maketitle

Phase transitions are among the most fundamental concepts of condensed matter physics. It is standard to classify equilibrium phase transitions as either first‑order or continuous. In particular, quantum phase transitions occur in the ground state as a control parameter is tuned~\cite{Sachdev2011}. At a first-order quantum phase transition, the two lowest energy levels cross linearly and exchange the identity of the ground state, leaving a cusp in its energy~\cite{Sachdev2011,Campostrini2014}. At a continuous quantum phase transition, they instead meet tangentially, and the gap closes smoothly as a power law. Moreover, the theory developed to explain the rich critical phenomena accompanying continuous phase transitions has become a central organizing principle of condensed‑matter physics.


Beyond equilibrium, quantum dynamics can give rise to new transitions that have no static counterpart~\cite{Dziarmaga2010, Polkovnikov2011, DAlessio2016, Mitra2018}. These dynamical phase transitions not only challenge the established paradigm of conventional phase transitions, posing fundamental theoretical issues in statistical physics~\cite{Chiocchetta2017,Jian2019,Madeira2024,Mittal2026,Heyl2013,Huang2019}, but also carry significant potential applications in fast-developing quantum devices and quantum circuits~\cite{Li2018,Li2019,Skinner2019,Fisher2023,Liu2024,Liu2024Noise, Liu2026Noisy, Zhang2026,Zhu2023}. The entanglement spectrum serves as a powerful diagnostic of quantum phases and criticality~\cite{Li2008,Pollmann2010,DeChiara2012,Canovi2014,Torlai2014,DeNicola2021,Gong2018,Potirniche2017,Potter2016,Lu2019,Jafari2021}.

\begin{figure}[t!]
  \includegraphics[width=0.95\columnwidth]{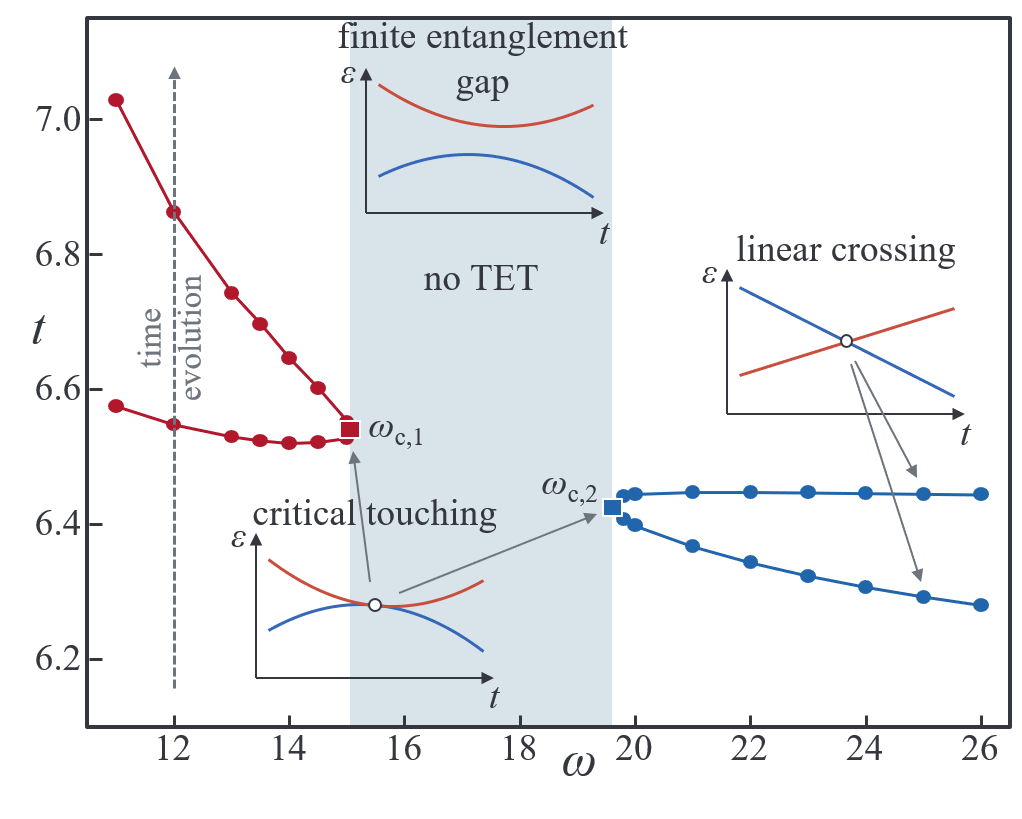}
\caption{Phase diagram of $J_2=-0.5$ and $h_0=4$ model on the $(\omega, t)$ plane. The two first-order TET branches terminate pairwise at the critical touchings, enclosing a finite window without crossings (shaded). Colored dots mark the first-order TETs, where the leading parity levels cross linearly and the dominant Schmidt parity switches. At critical frequencies, two branches of first-order TETs merge into critical touching (two squares), where the two lowest levels meet tangentially without a parity switch. In the shaded region between the two critical frequencies, the entanglement gap remains finite within the time window considered here and no TET occurs.}
  \label{fig:classification}
\end{figure}

The temporal entanglement transition (TET) is a newly identified nonanalyticity of the entanglement spectrum that develops during real-time evolution~\cite{Gadge2026}. It occurs when the two leading Schmidt levels in different symmetry sectors cross linearly in time, so that the dominant symmetry sector switches and the lowest entanglement energy develops a cusp. These transitions are marked by a simultaneous vanishing of the entanglement echo~\cite{Poyhonen2021} and the Schmidt gap, $\delta \lambda = \lambda_+ - \lambda_-$ between the two largest Schmidt values, but remain invisible to conventional observables such as the magnetization and the Loschmidt echo. All TETs reported so far follow the linear-crossing pattern of Schmidt levels as illustrated in the right inset of Fig.~\ref{fig:classification}, which can be viewed as a temporal analog of a first-order transition in equilibrium quantum phase transitions. This raises a natural question: can ``continuous'' TET, characterized by critical touching in the entanglement spectrum, emerge in nonequilibrium quantum dynamics? If so, what critical phenomena would accompany such a transition?

In this paper, we give a positive answer to these questions by reporting the discovery of a critical touching TET that exhibits a tangential contact of the leading parity-resolved Schmidt levels, a temporal analog of a continuous quantum phase transition in equilibrium. As illustrated in the lower-left inset of Fig.~\ref{fig:classification}, this critical touching arises from the merging of two linear crossings. At a fixed drive frequency, two linear crossings occur during the evolution, separating the time window into regions with opposite dominant parity. As the drive frequency is tuned, these two crossings approach each other and at a critical frequency they merge into a single critical touching, where the levels meet tangentially, without the cusp that characterizes a linear-crossing TET. Beyond this critical point, we find that the entanglement gap remains finite as shown Fig.~\ref{fig:classification}.

We identify two ingredients that bring this coalescence to a finite frequency. In the product-state limit, the two first-order TETs merge only as $\omega\rightarrow\infty$. A weak symmetry-preserving perturbation of the initial state rounds the product-state cusp, while finite-frequency Floquet corrections tune the resulting extremum of $\deltalambda$ through zero. 
We use the initial entropy to quantify the deviation from the product-state limit as the initial transverse field is varied.
As $\omega$ approaches $\omega_{c,1}$ from below, the temporal separation $W$ of the two first-order TETs vanishes as $W\propto(\omega_{c,1}-\omega)^{0.43(1)}$ and its derivative $dW/d\omega$ diverges, exhibiting a distinct nonanalytic signal absent in a linear-crossing TET. Then the gap between two largest Schmidt levels opens at some critical time. As $\omega$ increases, we further find a second critical frequency $\omega_{c,2}$, at which the Schmidt gap closes by a critical touching at the critical time. When $\omega>\omega_{c,2}$ the two linear-crossing TETs reappear. These features are captured by a second-order Floquet Hamiltonian and are robust across a broad range of coupling ratios, establishing the critical touching as a new class of TETs. Our results lay a new foundation for the classification of nonequilibrium entanglement phase transitions and bring novel insights to the research of fundamental entanglement phase transitions.

\textit{Model and method.---} We study a periodically driven transverse-field $J_1$--$J_2$ Ising chain described by Hamiltonian
\begin{align}
 H(t)=-J_1\sum_{i=1}^{L-1}S_i^zS_{i+1}^z -J_2\sum_{i=1}^{L-2}S_i^zS_{i+2}^z -h(t)\sum_{i=1}^{L}S_i^x,
 \label{eq:model}
\end{align}
where $S_i^\alpha=\sigma_i^\alpha/2$ denotes the spin operator on site $i$, and $J_1$ and $J_2$ are the nearest- and next-nearest-neighbor Ising couplings, respectively. The chain is driven by a periodic transverse field $h(t)=h_0\cos(\omega t)$. We impose open boundary conditions and take $J_1=1$ as the energy unit. At zero transverse field, the classical $J_1$--$J_2$ Ising chain has a multiphase point $J_2/J_1=-0.5$ separating the ferromagnetic and period-four $\uparrow\uparrow\downarrow\downarrow$ ground states~\cite{Selke1988}.

\begin{figure}[t]
  \includegraphics[width=\columnwidth]{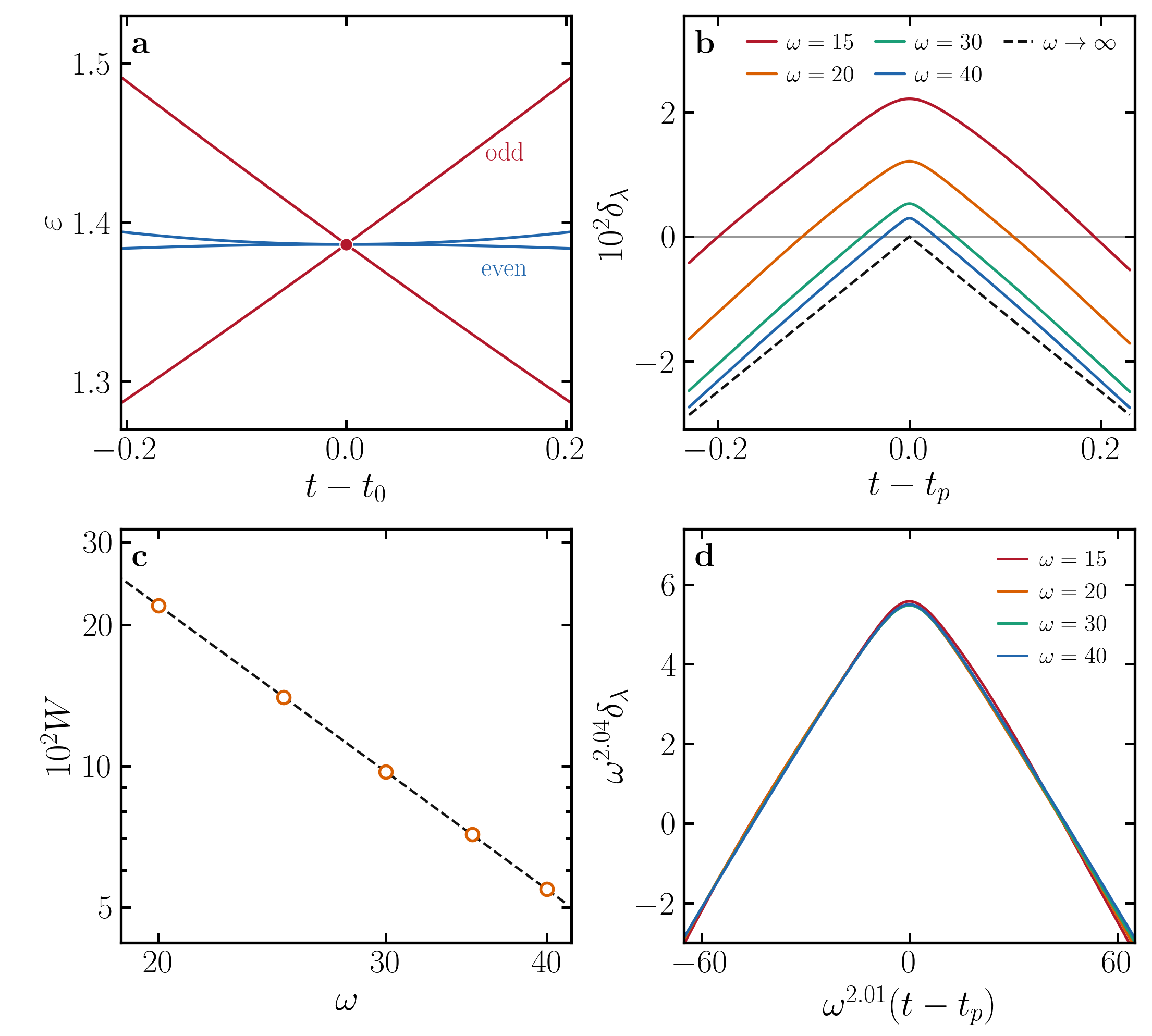}
\caption{Product-state limit at $J_2/J_1=-0.5$. (a) Entanglement energies near the fourfold degenerate time $t_0=\pi/|J_2|=2\pi$ in the $\omega\rightarrow\infty$ limit. The odd-parity levels (red) cross linearly while the even-parity levels (blue) remain degenerate at linear order. (b) Temporal entanglement gap $\deltalambda$ at finite drive frequencies. The cusp in the $\omega\rightarrow\infty$ result (black dashed line) resolves into a positive peak at $t_p$ bounded by two parity-switching crossings $t_1$ and $t_2$. (c) Crossing separation $W=t_2-t_1$ versus frequency on log-log scales. The dashed line presents a power-law dependence with exponent $2.01$. (d) The data collapsing of the $\deltalambda$ profiles exhibiting a universal scaling law.}
  \label{fig:product}
\end{figure}

The entanglement dynamics of the system is investigated through the time-dependent variational principle (TDVP) for matrix-product states ~\cite{Haegeman2011, Haegeman2016, Schollwock2011}. We divide the chain into a subsystem $A$ of the first $L_A$ sites and its complement $\bar A$ and calculate the reduced density matrix $\rho_A(t)=\operatorname{Tr}_{\bar A}|\psi(t)\rangle\langle\psi(t)|$ and its eigenvalues $\lambda_i(t)$ (Schmidt values) at each instant of time $t$. The entanglement Hamiltonian is defined as $H_E(t)=-\ln\rho_A(t)$, whose eigenvalues $\epsilon_i(t)=-\ln\lambda_i(t)$ form the entanglement spectrum and $\epsilon_0(t)=-\ln\lambda_{\max}(t)$ gives the lowest entanglement energy. 
The R\'enyi entropies $S_n(\rho_A)=\ln(\sum_i\lambda_i^n)/(1-n)$ probe $H_E$ at fictitious temperatures $T=1/n$, as $\operatorname{Tr}\rho_A^n=\operatorname{Tr}e^{-nH_E}$ is its partition function. In most of the calculations, we use $L=40$, $L_A=10$ and TDVP bond dimension $\chi=200$ to obtain highly accurate results. The numerical details are provided in the Supplemental Material.

The driven Hamiltonian and the initial states used in our main calculations preserve the global Ising parity $\mathcal P=\prod_{i=1}^{L}\sigma_i^x$. The reduced density matrix $\rho_A(t)$ therefore commutes with the subsystem parity $\mathcal P_A=\prod_{i\in A}\sigma_i^x$, and each Schmidt state carries a parity label $p=\pm1$~\cite{Goldstein2018}. We denote the largest Schmidt value in each sector by $\lambda_p(t)$ and track their difference $\deltalambda(t,\omega)=\lambda_+(t,\omega)-\lambda_-(t,\omega)$. In the temporal windows considered below, $\lambda_+$ and $\lambda_-$ are the two largest eigenvalues of $\rho_A$, and the Schmidt gap~\cite{DeChiara2012} is therefore $|\deltalambda|$. In the previous study~\cite{Gadge2026}, $\lambda_+$ and $\lambda_-$ cross linearly at a TET, leading to a vanishing of $\deltalambda$, a cusp in $\epsilon_0(t)$ and a parity switching of the dominant Schmidt state. The crossing of the two lowest levels of $H_E(t)$ can be viewed as a temporal analog of a first-order quantum phase transition in equilibrium.

\textit{Anomalous linear touching in product-state limit.---} We first consider the fully $x$-polarized product state $|{+x}\rangle^{\otimes L}$ as the initial state. In the high-frequency limit, the transverse driving field averages to zero and the evolution is governed by the Ising Hamiltonian
\begin{equation}
 H_I=-J_1\sum_{i=1}^{L-1}S_i^zS_{i+1}^z - J_2\sum_{i=1}^{L-2}S_i^zS_{i+2}^z.
 \label{eq:hi}
\end{equation}
Since all terms in $H_I$ commute, only the interaction gates connecting the two subsystems contribute to the Schmidt values. At $t_0=\frac{\pi}{|J_2|}$ the two $J_2$ gates each accumulate a phase $|J_2|t_0/4=\pi/4$. The four-dimensional boundary subspace is then maximally mixed with four degenerate Schmidt values $\lambda=1/4$, independent of the dynamical phase from the $J_1$ gate. The detailed derivation is provided in the Supplemental Material.

Away from $t_0$, the four degenerate Schmidt values split linearly in $\tau=t-t_0$ as $\lambda_{e,\pm}=\frac14 \pm \Big|\frac{J_2 \tau}{4} \cos\frac{\pi J_1}{4J_2}\Big|+O(\tau^2)$ and $\lambda_{o,\pm}=\frac14 \pm \Big|\frac{J_2 \tau}{4} \sin\frac{\pi J_1}{4J_2}\Big|+O(\tau^2)$ with ``o'' and ``e'' denoting the odd and even parities, respectively.
Accordingly, this leads to a linear closing of Schmidt gap $\deltalambda = (|\cos\frac{\pi J_1}{4J_2}|-|\sin\frac{\pi J_1}{4J_2}|)\frac{|J_2 \tau|}{4}$ between $\lambda_{e,+}$ and $\lambda_{o,+}$, as illustrated in Fig.~\ref{fig:product}(a).

\begin{figure}[t]
  \includegraphics[width=\columnwidth]{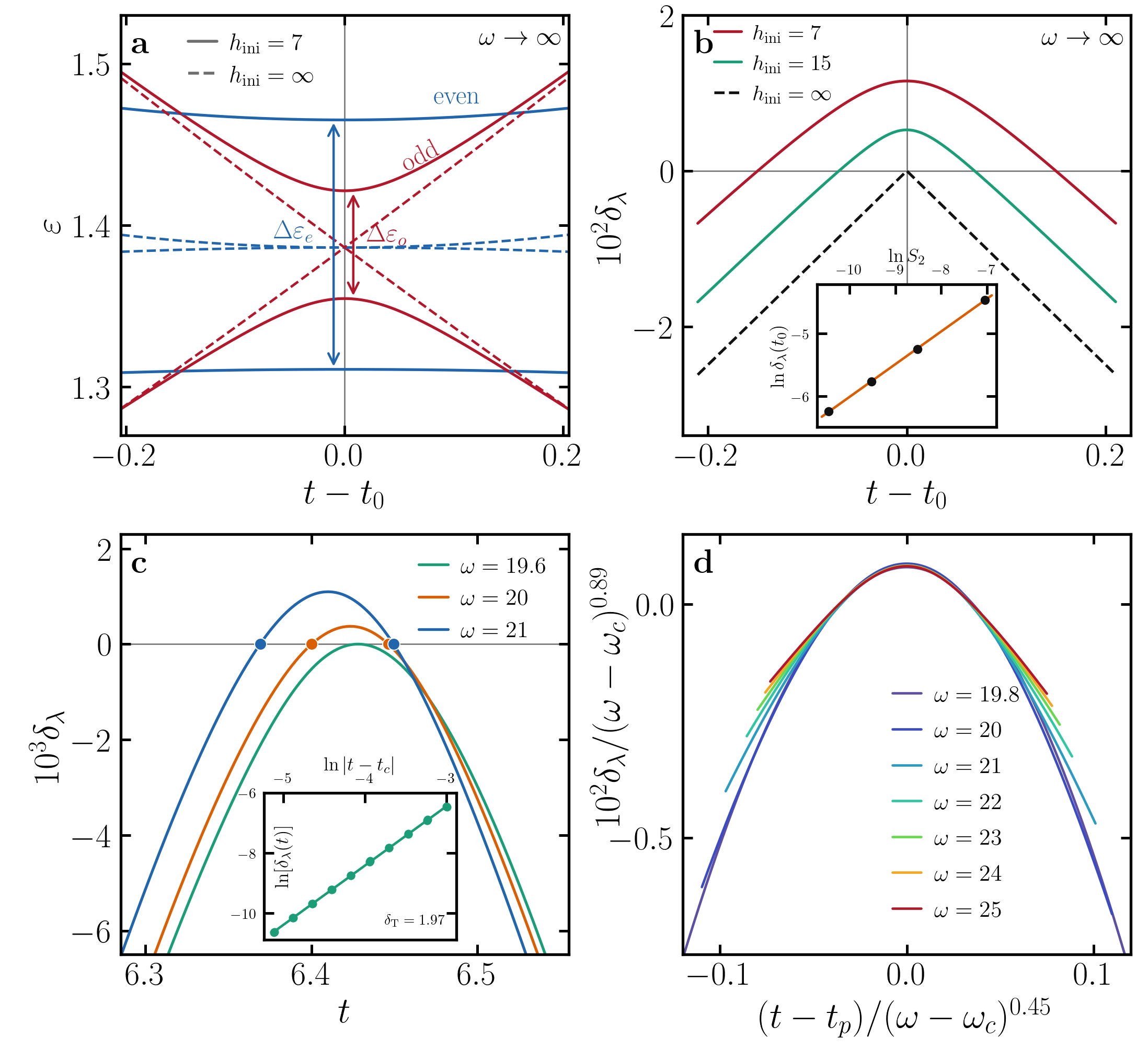}
\caption{Critical touching from a weakly perturbed product state. (a) The two lowest entanglement energies in each parity sector for the product state (dashed lines) and for the $\hini/J_1=7$ ground state (solid lines). The initial-state perturbation opens unequal gaps $\Delta\varepsilon_o$ and $\Delta\varepsilon_e$, producing two linear crossings between the leading levels of opposite parity. (b) Evolution under the commuting Hamiltonian $H_I$ for different initial transverse fields. Each rounded peak is bounded by two first-order TETs, whose separation increases as $\hini$ decreases. Inset: log plot of the peak height $\deltalambda(t_0)$ against the initial R\'enyi entropy $S_2(0)$. (c) Evolution of the driven Hamiltonian $H(t)$ near the critical frequency $\omega_{c,2}=19.61$. Inset: log plot of $|\deltalambda(t)|$ versus $|t-t_c|$ at $\omega_{c,2}$. (d) Data collapsing of the peak profiles $\deltalambda(t)$ for $\omega$ from $19.8$ to $25$, with the two axes rescaled by the independently measured powers $(\omega-\omega_{c,2})^{0.45}$ and $(\omega-\omega_{c,2})^{0.89}$, respectively.}
  \label{fig:rounding}
\end{figure}

It is worth noting that the linear closing of $\deltalambda$ at $t_0$ is remarkably different from the linear-crossing TETs reported in Ref.~\cite{Gadge2026}, since it does not involve the changing of the dominant Schmidt parity. The full driven evolution of $H(t)$ at finite frequency reveals that the origin of this anomalous closing is the coalescence of two first-order TETs at infinite drive frequency. To see this coalescence, we evolve the $x$-polarized state with $J_2=-0.5$ and $h_0=4$ over a range of frequencies $\omega$. This $J_2/J_1$ ratio is chosen for simplicity, since the $\lambda_{o,\pm}$ pair splits maximally while the $\lambda_{e,\pm}$ pair remains at $1/4$. The phenomena below occur in a broad parameter range. At finite frequency, the cusp resolves into a positive peak at $t_p$ bounded by two parity-changing crossings, representing two first-order TETs occurring at $t_1$ and $t_2$ as shown in Fig.~\ref{fig:product}(b). With increasing frequency, the two crossings approach each other, and their separation vanishes as $W=t_2-t_1\propto\omega^{-2.01}$, as illustrated in Fig.~\ref{fig:product}(c). The two TETs therefore coalesce as $\omega\rightarrow\infty$, explaining the absence of the parity switching in the anomalous closing.

Moreover, we find that this coalescence of the two first-order TETs can well be described by a universal scaling form
\begin{equation}
\deltalambda(t,\omega)=\omega^{\Delta_{A}}F_1(\omega^{-\beta_{A}}(t-t_p)).
\label{infiome}
\end{equation}
The fitting exponents, $\Delta_{A}=-2.04$ and $\beta_{A}=-2.01$, are both close to 2, suggesting that this coalescence of TETs is governed by a $1/\omega^2$ scale in high-frequency limit. Eq.~(\ref{infiome}) is verified by the scaling collapsing depicted in Fig.~\ref{fig:product}(d).

\textit{Finite-frequency critical touching from a perturbed product state.---} In the product-state limit, the two first-order TETs coalesce only at infinite drive frequency. It is natural to ask whether this coalescence can be brought down to a finite frequency, producing a new type of TET. Here we show that a weak symmetry-preserving perturbation of the product state brings the coalescence to a finite critical frequency. We generate a controlled family of initial states as the ground states of $H_{\mathrm{ini}}=H_I-\hini\sum_iS_i^x$. The initial state approaches $|{+x}\rangle^{\otimes L}$ as $\hini\rightarrow\infty$. For large but finite $\hini$, lowering $\hini$ moves the state away from this limit while preserving Ising parity. For these initial states, we use the second R\'enyi entropy $S_2(0)=-\ln\operatorname{Tr}\rho_A^2(0)$ to quantify their deviation from the product-state limit; for example, $S_2(0)=8.7\times10^{-4}$ at $\hini=7$.

In the $\omega\rightarrow\infty$ limit, the initial-state perturbation lifts the intra-parity degeneracies and rounds the product-state cusp into a smooth local extremum [Fig.~\ref{fig:rounding}(a)]. The odd- and even-parity gaps open unequally, $\Delta\epsilon_o<\Delta\epsilon_e$, so the leading branches cross once on each side of $t_0$. These crossings are first-order TETs, and their separation increases as $\hini$ decreases [Fig.~\ref{fig:rounding}(b)]. 
For $J_{2,\mathrm{ini}}=J_2=-0.5J_1$, varying $\hini$ gives $\deltalambda(t_0)\propto S_2(0)^{0.52}$ under evolution with $H_I$, as shown in the inset of Fig.~\ref{fig:rounding}(b). The near-square-root dependence is consistent with the quadratic perturbative onset of the second R\'enyi entropy about a product state~\cite{Zanardi2013}. However, the initial entropy alone does not determine the subsequent entanglement dynamics. Different perturbed initial states with the same $S_2(0)$ can have different critical frequencies (see SM).


Lowering the drive frequency from infinity moves the two TETs toward each other until they merge at $\omega_{c,2}=19.61$, as shown in Fig.~\ref{fig:rounding}(c). The properties of the level touching at $\omega_{c,2}$ is characterized by the gap profile $\deltalambda(t)$ around touching point $t_c$. In the inset of Fig.~\ref{fig:rounding}(c), we show that the gap profile can be fitted by a power-law function $\deltalambda(t)\propto|t-t_c|^{{\delta_{\mathrm T}}}$ with the temporal touching exponent ${\delta_{\mathrm T}=1.97(1)}$, indicating that the touching is tangential, $\partial_t\deltalambda(t_c,\omega_{c,2})=0$, rather than a linear crossing. The dominant parity does not exchange at touching point $t_c$, in contrast to a first-order TET where the linear crossing switches the parity.

As $\omega$ approaches $\omega_{c,2}$, both the temporal separation $W$ and the peak height $\deltalambda(t_p)$ vanish with power-law behavior. We characterize these decays by $W\propto(\omega-\omega_{c,2})^{\beta_{\mathrm T}}$, with the temporal order-parameter exponent $\beta_{\mathrm T}=0.45(1)$, and $\deltalambda(t_p)\propto(\omega-\omega_{c,2})^{\Delta_{\mathrm T}}$, with the temporal gap exponent $\Delta_{\mathrm T}=0.89(1)$, as obtained from independent fits (see SM). Moreover, we discover that near $\omega_{c,2}$, $\deltalambda$ satisfies a universal scaling form
\begin{equation}
\deltalambda(t) = (\omega-\omega_{c,2})^{\Delta_{\mathrm T}}
F_2[(\omega-\omega_{c,2})^{-\beta_{\mathrm T}}(t-t_p)],
\label{eq:scaling}
\end{equation}
{as confirmed by the data collapse in Fig.~\ref{fig:rounding}(d). The independently fitted exponents satisfy the scaling law $\Delta_{\mathrm T}=\beta_{\mathrm T}\delta_{\mathrm T}$ within numerical accuracy, linking the scaling of the peak height, crossing separation, and local touching profile. A derivation of this relation is given in the Supplemental Material.} Interestingly, the critical exponents here are remarkably distinct from those in Eq.~(\ref{infiome}), indicating that the finite initial entanglement can give rise to different universality class of TETs. {In addition, $dW/d\omega\propto(\omega-\omega_{c,2})^{\beta_{\mathrm T}-1}$, with $\beta_{\mathrm T}-1=-0.55(1)$, diverges at $\omega_{c,2}$ and provides a nonanalytic signal of the merging.}

This critical touching behavior presents a temporal analog of a equilibrium continuous phase transition at the liquid-gas critical point~\cite{Widom1965}, where the separation $W$ plays the role of the density difference of the two branches. {Its temporal order-parameter exponent $\beta_{\mathrm T}=0.45(1)$ lies below the mean-field liquid--gas value $1/2$, indicating a departure from the mean-field square-root scaling.}

Extending the frequency scan to lower frequencies reveals another critical touching occured at $\omega_{c,1}\simeq 15.05$, with the previous one denoted $\omega_{c,2}=19.61$. As shown in the phase diagram in Fig.~\ref{fig:classification}, increasing $\omega$ through $\omega_{c,1}$ brings two first-order TETs together and removes them, and the crossings remain absent until they reappear at $\omega_{c,2}$. The two critical touchings therefore bound a finite window in which no TET occurs. The critical touching at $\omega_{c,1}$ also shares the local structure of the liquid-gas critical endpoint. Moreover, we confirm that the scaling form of Eq.~(\ref{eq:scaling}) remains applicable, demonstrating the universality of the scaling properties. This second critical TETs reappeared at $\omega_{c,2}=19.61$ phase is clearly beyond the liquid-gas case, where the coexistence region closes permanently at the sole endpoint and never reopens.

Our further calculations show that the critical touching occurs in a broad range of $J_2/J_1$ ratios, e.g. $-0.8$, $-0.5$, and $-0.3$. While the critical frequency and detailed forms of the Schmidt levels $\lambda$ vary appreciably with $J_2/J_1$, the scaling behaviors around the critical frequency remains intact in all cases studied. Breaking Ising parity in the initial state turns both linear TETs and critical touchings into avoided crossings (see SM).

\textit{Floquet effective Hamiltonian.---} The phenomena reported above persist up to high drive frequencies, indicating that the underlying physics can be understood through an effective Hamiltonian. According to Floquet-Magnus theory~\cite{Bukov2015,Eckardt2017,Eckardt2015,Goldman2014}, the effective Hamiltonian to second order in the Floquet gauge takes the form
\begin{align}
 &H_F=-\sum_{r=1,2}J_r\sum_{i=1}^{L-r}
 \left(1-\frac{h_0^2}{2\omega^2}\right)S_i^zS_{i+r}^z\nonumber\\
 &-\frac{h_0^2}{2\omega^2}\sum_{r=1,2}J_r
 \sum_{i=1}^{L-r}S_i^yS_{i+r}^y + \frac{h_0}{\omega^2}\sum_{i=1}^{L}S_i^x B_i^2+O(\omega^{-3}),
 \label{eq:hfexpanded}
\end{align}
where $B_i=\sum_{\substack{r=1,2;\ s=\pm1}} J_rS_{i+sr}^z$. The first term renormalizes the Ising couplings, while the rest are $\omega^{-2}$ order terms containing $YY$ interactions, transverse fields, and three-spin $S^zS^xS^z$ interactions, which preserve the global parity but break the commuting structure of $H_I$. Without the $\omega^{-2}$ terms, the static Hamiltonian $H_I$ alone gives two first-order TETs at every frequency. The critical touching and the no-TET window are reproduced only when the $\omega^{-2}$ terms are included. As presented in SM, the frequency scan generated by effective Hamiltonian $H_F$ exhibits both the disappearance and the reappearance of the two crossings with critical frequencies $\omega_{c,1}^{(F)}=14.71$ and $\omega_{c,2}^{(F)}=20.28$, in reasonable agreement with the full driven evolution.

\textit{Discussion.---} Our results extend the classification of temporal entanglement transitions beyond linear crossings. At a first-order TET, the two leading parity-resolved Schmidt levels cross linearly, producing a cusp in the lowest entanglement energy. The transition found here instead occurs through the coalescence of two such crossings. At the critical frequency, the levels meet tangentially without exchanging the dominant parity, and the lowest entanglement energy remains smooth. As the frequency approaches this point, the two first-order transition times merge continuously, with their separation $W=t_2-t_1$ vanishing at $\omega_c$.

The two scaling collapses provide a unified quantitative description of this change. Near the finite-frequency endpoint, with $r=\omega-\omega_{c,2}$, the scaling form is $\deltalambda=r^{\Delta_{\mathrm T}} F[(t-t_p)/r^{\beta_{\mathrm T}}]$. The independently measured exponents $\beta_{\mathrm T}=0.45(1)$, $\Delta_{\mathrm T}=0.89(1)$, and $\delta_{\mathrm T}=1.97(1)$ satisfy $\Delta_{\mathrm T}=\beta_{\mathrm T}\delta_{\mathrm T}$ within numerical accuracy. This relation shows that the peak-height scaling is not independent. It is fixed by the temporal separation of the two crossings and the local order of their touching. The scaling collapse therefore connects the horizontal contraction of the two crossing times with the vertical closing of the Schmidt gap. In the product-state limit, by contrast, the corresponding ratio $\Delta_{A}/\beta_{A}\simeq1.01$ is consistent with the exact linear cusp. The two limits also differ in their frequency dependence: the product-state coalescence is controlled by the $1/\omega^2$ high-frequency scale, whereas the finite-frequency endpoint has $\beta_{\mathrm T}$ below the mean-field value $1/2$. The derivative $dW/d\omega\propto r^{\beta_{\mathrm T}-1}$ diverges at the finite endpoint and gives the corresponding singular response of the crossing separation to frequency tuning.

Two effects produce the finite-frequency critical touching. A symmetry-preserving perturbation of the product initial state lifts the intra-parity degeneracies and rounds the product-state cusp. Finite-frequency Floquet corrections then tune the rounded extremum of $\deltalambda$ through zero and determine the critical frequencies. 
The initial entropy provides a measure of deviation from the product-state limit as $\hini$ is varied. States prepared by different Hamiltonians can have the same initial entropy but different critical frequencies (see SM). 
The fact that $H_F$, but not $H_I$, reproduces the disappearance and reappearance of the crossings, together with their persistence across different $J_2/J_1$, shows that the mechanism is neither a time-averaged effect nor a feature unique to the equilibrium multiphase point.

As the drive frequency increases, the two branches of first-order TETs disappear at $\omega_{c,1}$ and reappear at $\omega_{c,2}$. Each critical touching locally resembles a liquid--gas endpoint, but the resulting reentrant structure has no direct equilibrium counterpart. More broadly, our results show that first-order and continuous TETs can occur within the same dynamical phase diagram. It will be interesting to extend the critical touching picture to different driven systems and symmetry classes.

{\it Acknowledgments: } YFJ is supported in part by the National Key R\&D Program of China under Grants No. 2022YFA1402703, NSFC under Grant No. 12347107 and 12574160. SY is supported by the National Natural Science Foundation of China (Grants No. 12675053 and 12222515), the Research Center for Magnetoelectric Physics of Guangdong Province (Grant No. 2024B0303390001), the Guangdong Provincial Key Laboratory of Magnetoelectric Physics and Devices (Grant No. 2022B1212010008), and Quantum Science and Technology-National Science and Technology Major Project(Grant No.2025ZD0300400). SXZ is supported by the National Natural Science Foundation of China (No. 12574546), Quantum Science and Technology-National Science and Technology Major Project (No. 2024ZD0301700), and the Chinese Academy of Sciences (No. XDB1680201 and No. YSBR-150).

\bibliography{main}

\clearpage
\onecolumngrid
\raggedbottom
\setcounter{secnumdepth}{2}
\setcounter{section}{0}
\setcounter{subsection}{0}
\setcounter{figure}{0}
\setcounter{table}{0}
\setcounter{equation}{0}
\renewcommand{\thefigure}{S\arabic{figure}}
\renewcommand{\thetable}{S\arabic{table}}
\renewcommand{\theequation}{S\arabic{equation}}
\renewcommand{\theHfigure}{supp.figure.\arabic{figure}}
\renewcommand{\theHtable}{supp.table.\arabic{table}}
\renewcommand{\theHequation}{supp.equation.\arabic{equation}}
\renewcommand{\theHsection}{supp.section.\arabic{section}}
\renewcommand{\theHsubsection}{supp.subsection.\arabic{section}.\arabic{subsection}}

\begin{center}
{\large\bfseries Supplemental Material for ``Critical Touching of Temporal Entanglement Transitions''}
\end{center}
\vspace{0.5em}

\section{Numerical methods and conventions}
\label{sec:methods}

The principal calculations in this study use $J_1=1$, $J_2=J_{2,\mathrm{ini}}=-0.5$, $h_0=4$, and $\hini=7$; parameters for additional tests are specified where they are used. All chains have open boundaries, with subsystem $A$ consisting of the first $L_A$ sites. The MPS calculations output Schmidt amplitudes $\sqrt{\lambda_i}$, whereas all reported $\lambda_i$ are eigenvalues of $\rho_A$. The parity-resolved levels and characteristic times use the notation introduced in the main text.

The $L=40$ calculations use GPU-based two-site time-dependent variational-principle (TDVP) evolution of matrix-product states~\cite{Haegeman2011,Haegeman2016,Schollwock2011}, with $L_A=10$. At each time step $t_n=n\Delta t$, the code evaluates the transverse field $h(t_n)=h_0\cos(\omega t_n)$, rebuilds the Hamiltonian matrix-product operator, and performs one left-to-right and one right-to-left TDVP sweep. The local effective evolutions are evaluated by Krylov exponentiation with a stopping threshold of $10^{-8}$.
The finite-$\hini$ initial states are obtained by DMRG, while $|+x\rangle^{\otimes L}$ is represented exactly by a bond-dimension-one matrix-product state. The principal finite-$\hini$ runs use $\chi=200$ and $\Delta t=0.005$.

\FloatBarrier
\section{Product-state limit}

For $\hini=\infty$, the initial state is $|+x\rangle^{\otimes L}$. Under the Ising Hamiltonian
\begin{equation}
H_I=-J_1\sum_iS_i^zS_{i+1}^z-J_2\sum_iS_i^zS_{i+2}^z,
\end{equation}
since all Ising terms commute, the internal Ising terms within $A$ or $\bar A$ are local unitaries with respect to the bipartition and do not change the Schmidt values. For this boundary calculation, label the four spins adjacent to the cut as sites $1,2,3,4$, with the cut between sites $2$ and $3$. Sites $1,2$ belong to $A$, while sites $3,4$ belong to $\bar A$. After removing the internal local unitaries, the Schmidt spectrum is determined by the following four-spin state:
\begin{equation*}
 |\psi_\partial(t)\rangle=
 \sum_{z_1,z_2,z_3,z_4=\pm1}
 C_{(z_1,z_2),(z_3,z_4)}(t)
 |z_1,z_2\rangle_A|z_3,z_4\rangle_{\bar A},
\end{equation*}
where $z_j=\pm1$ is the eigenvalue of $\sigma_j^z=2S_j^z$. Thus $C$ is a $4\times4$ matrix of wave-function amplitudes: 
\begin{equation}
 C_{(z_1,z_2),(z_3,z_4)}=\frac14
 \exp\!\left\{\frac{it}{4}\left[J_1z_2z_3
 +J_2z_1z_3+J_2z_2z_4\right]\right\}.
 \label{eq:boundarymatrix}
\end{equation}
The factor $1/4$ comes from the equal amplitudes of the four spins in the initial $|+x\rangle$ state. Tracing out the two boundary spins in $\bar A$ gives $\rho_{A,\partial}=CC^\dagger$, which has the same nonzero eigenvalues as $\rho_A$. In the basis $\boldsymbol z=(z_1,z_2)$, the matrix elements of $\rho_{A,\partial}$ are
\begin{align}
 (\rho_{A,\partial})_{\boldsymbol z,\boldsymbol z'}={}&\frac14
 \cos\!\left\{\frac{t}{4}
 \left[J_1\Delta z_2+J_2\Delta z_1\right]\right\}
 \nonumber\\
 &\times\cos\!\left(\frac{J_2t}{4}\Delta z_2\right),
 \label{eq:boundaryrho}
\end{align}
where $\Delta z_j=z_j-z_j'=0,\pm2$. At the sequence of times
\begin{equation}
 t_m=\frac{(2m+1)\pi}{|J_2|},\qquad m=0,1,\ldots,
 \label{eq:generalclock}
\end{equation}
every off-diagonal element in Eq.~\eqref{eq:boundaryrho} vanishes, and $\rho_{A,\partial}(t_m)$ is proportional to the identity matrix, $\frac14 I_4$.
In the boundary representation, $\mathcal P_A$ acts as $\mathcal P_\partial=\sigma_1^x\sigma_2^x$, which maps $|z_1,z_2\rangle$ to $|-z_1,-z_2\rangle$. The symmetric and antisymmetric combinations of each paired configuration have even and odd parity, respectively, giving two states in each sector. Since the internal Ising unitaries commute with $\mathcal P_A$, each parity sector contains two nonzero Schmidt values, which gives total sector weights $w_+=w_-=1/2$.

The local splitting can be derived by expanding Eq.~\eqref{eq:boundaryrho} in $\tau=t-t_0$. The diagonal elements remain $1/4$. The off-diagonal matrix elements for flipping $z_1$, $z_2$, or both spins are, respectively, $-|J_2|\tau/8$, $-(|J_2|\tau/8)\cos(\pi\eta/2)$ and $-(|J_2|\tau/8)\cos[\pi(\eta\mp1)/2]$ to leading order, where $\eta=J_1/|J_2|$. Diagonalizing the two parity blocks gives shifts of magnitude $(|J_2||\tau|/4)|\cos(\pi\eta/4)|$ in the even sector and $(|J_2||\tau|/4)|\sin(\pi\eta/4)|$ in the odd sector, as quoted in the main text. For $J_2=-0.5$ ($\eta=2$), $t_0=2\pi$ and the four levels are
\begin{equation}
{
\begin{aligned}
 \lambda_{-,1}&=\frac14+\frac{|\tau|}{8}+\frac{\tau^2}{64}+O(|\tau|^3),&
 \lambda_{-,2}&=\frac14-\frac{|\tau|}{8}+\frac{\tau^2}{64}+O(|\tau|^3),\\
 \lambda_{+,1}&=\frac14+\frac{\tau^2}{64}+O(\tau^4),&
 \lambda_{+,2}&=\frac14-\frac{3\tau^2}{64}+O(\tau^4).
\end{aligned}}
\end{equation}
Thus $\deltalambda=-|\tau|/8+O(|\tau|^3)$. The dominant parity on the two sides of $t_0$ is the same, which distinguishes this product-state cusp from a parity-switching linear TET.

\subsection{Finite-frequency splitting of the product-state cusp}

\begin{figure}[!htbp]
  \centering
  \includegraphics[width=0.42\textwidth]{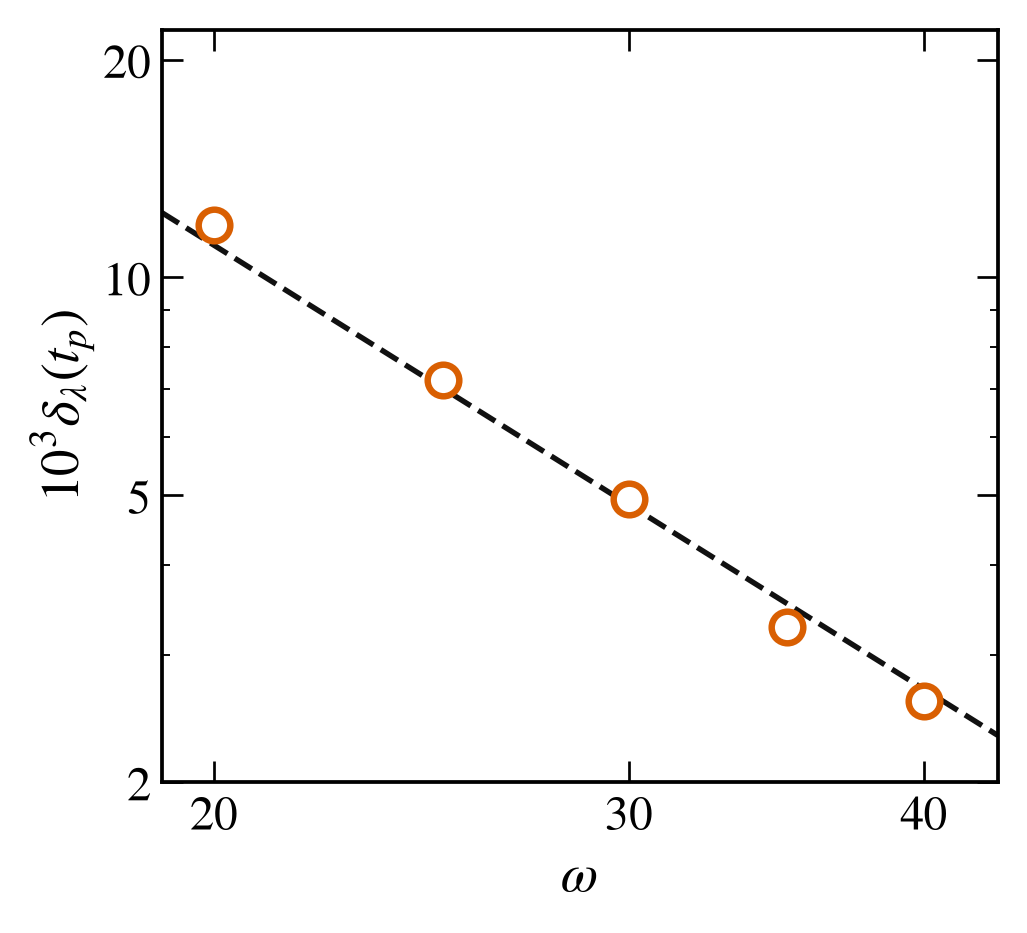}
\caption{Product-state peak-height scaling for $L=40$, $L_A=10$, $J_2=-0.5$, and $h_0=4$. Symbols are the TDVP peaks at $\omega=20,25,30,35,40$, obtained with $\chi=100$ and $\Delta t=0.02$. The dashed line shows the $\deltalambda(t_p)\propto\omega^{-2.04}$ dependence used for the vertical rescaling in Fig.~2(d) of the main text.}
  \label{fig:sm_product_peak}
\end{figure}

At finite frequency, the product-state cusp resolves into two parity-switching crossings enclosing a positive peak. For the $L=40$ TDVP data, the peak height and crossing separation scale as $\delta_\lambda(t_p)\propto\omega^{-2.04}$ and $W\propto\omega^{-2.01}$, respectively. These vertical and horizontal powers collapse the $\omega=15,20,30,40$ profiles in Fig.~2(d). Figure~\ref{fig:sm_product_peak} shows the peak-height data and the $\omega^{-2.04}$ dependence. Here the scaling variable is the drive frequency itself, and the crossings merge in the $\omega\to\infty$ limit.

\FloatBarrier
\section{Initial-state perturbations}

\subsection{Parity-preserving initial states}

We prepare the ground states of $H_{\mathrm{ini}}=H_I-\hini\sum_iS_i^x$, with $J_{2,\mathrm{ini}}=-0.5$ unless stated otherwise, and evolve them under $H_I$ to isolate the effect of the initial state. Reducing $\hini$ from infinity limit introduces weak entanglement while preserving Ising parity. We use the initial second R\'enyi entropy $S_2(0)=-\ln\operatorname{Tr}\rho_A^2(0)$ to quantify the deviation from the product-state limit. For $L=14$ and $L_A=7$, fits over $7\leq\hini\leq100$ give the odd- and even-sector gaps at $t_0=2\pi$ as approximately $2.19\,S_2(0)^{0.496}$ and $5.29\,S_2(0)^{0.502}$, respectively.
The odd-sector gap is smaller than the even-sector gap. Consequently, rounding the product-state degeneracy produces a positive leading-level difference at $t_0$, bounded by two linear TETs, as shown in Fig.~3(a),(b) of the main text. A power-law fit of the peak over $7\leq\hini\leq100$ gives
\begin{equation}
 \deltalambda(t_0)=0.435\,S_2(0)^{0.517(2)}.
\end{equation}
This approximately square-root dependence is consistent with perturbation about a product state: the subleading Schmidt amplitudes are first order in $\hini^{-1}$, while $S_2(0)$ is quadratic in those amplitudes~\cite{Zanardi2013}.

\subsection{Dependence on the initial Hamiltonian}
\label{sec:initial_hamiltonian}

\begin{figure}[!htbp]
\centering
\includegraphics[width=0.47\textwidth]{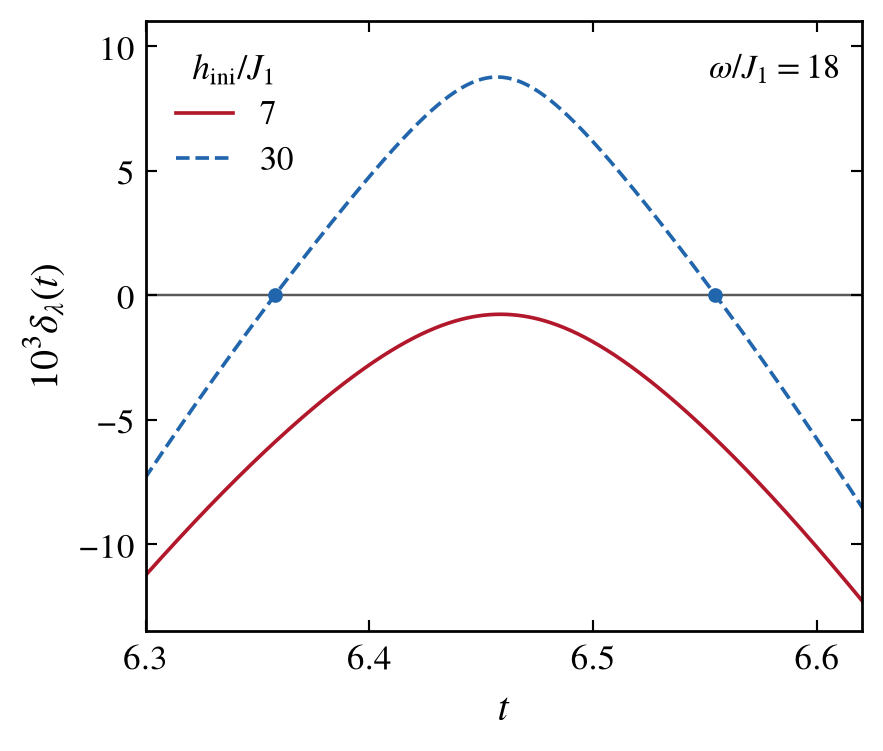}
\caption{Temporal entanglement dynamics from different initial states with same $S_2(0)$, for $L=14$, $L_A=7$, $J_2=-0.5$, $h_0=4$ and $\omega=18$. The red solid and blue dashed curves use the $\hini=7$ and $30$ initial states, respectively. The latter exhibits two parity-switching crossings, while the former remains gapped in the displayed window.}
\label{fig:sm_initial_hamiltonian}
\end{figure}

To test whether the initial entropy alone determines the subsequent dynamics, we compare parity-preserving ground states prepared at fixed $J_{1,\mathrm{ini}}=1$ with $(\hini,J_{2,\mathrm{ini}})=(7,-0.3)$ and $(30,-3.112)$, chosen so that the initial second R\'enyi entropies are approximately matched at $S_2(0)\simeq7.02\times10^{-4}$. We evolve both states under the same driven Hamiltonian with $J_2=-0.5$, $h_0=4$ and $\omega=18$. As shown in Fig.~\ref{fig:sm_initial_hamiltonian}, $\deltalambda(t)$ remains negative for $\hini=7$ in the displayed time window, whereas the $\hini=30$ state exhibits two parity-switching crossings. Thus, the initial entropy does not solely determine the subsequent crossing structure.

\FloatBarrier
\subsection{Effect of initial-state parity breaking}

\begin{figure}[!htbp]
\centering
\includegraphics[width=0.88\textwidth]{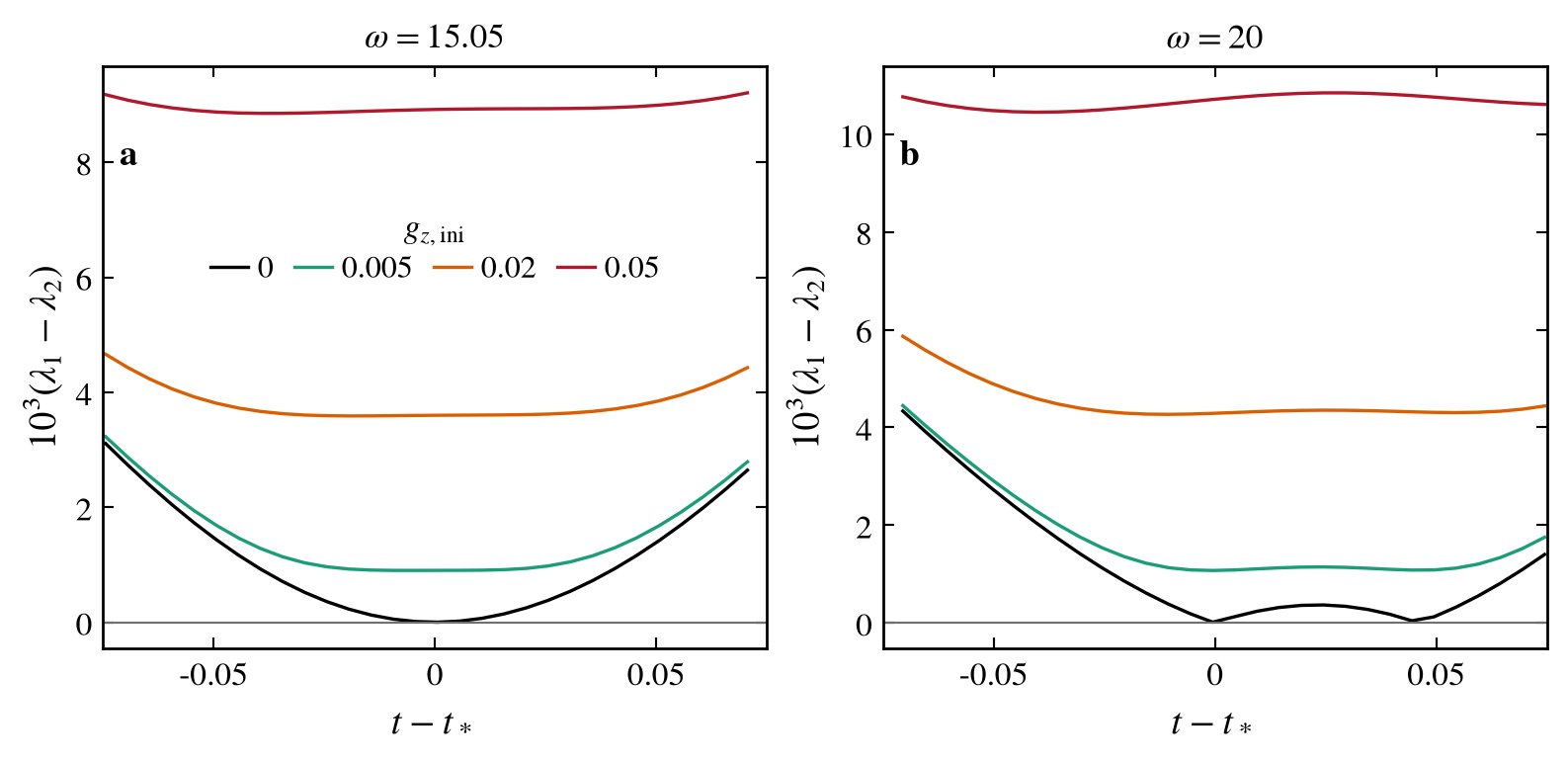}
\caption{Effect of initial-state parity breaking on the local Schmidt gap, at $L=14$, $L_A=7$, $J_2=-0.5$, $h_0=4$ and $\hini=7$. (a) Near the lower continuous TET of the parity-preserving preparation, $\omega=15.05$, with $t_*=6.54$. (b) Near the two linear TETs at $\omega=20$, with time centered on the first crossing, $t_*=6.40$. The legend gives the longitudinal fields applied only during preparation; $g_{z,\mathrm{ini}}=0$ is the parity-preserving reference.}
\label{fig:sm_z2_control}
\end{figure}

Here, we study the effect of parity breaking perturbation to the initial state by adding a longitudinal field only to the preparation Hamiltonian,
\begin{equation}
H_{\mathrm{ini}}(g_{z,\mathrm{ini}})=H_I-\hini\sum_iS_i^x-g_{z,\mathrm{ini}}\sum_iS_i^z,
\end{equation}
with $L=14$, $L_A=7$ and $\hini=7$. The evolution Hamiltonian remains $H(t)$ with $h_0=4$ and preserves Ising parity. Since the initial state is not a parity eigenstate for $g_{z,\mathrm{ini}}\neq0$, we track the full Schmidt gap $\lambda_1-\lambda_2$ defined by the two largest eigenvalues of $\rho_A$.
Figure~\ref{fig:sm_z2_control} compares the parity-preserving preparation at $g_{z,\mathrm{ini}}=0$ with initial states prepared at nonzero longitudinal fields. For nonzero $g_{z,\mathrm{ini}}$, a finite Schmidt gap replaces the gap closings at both the lower continuous TET and the two linear TETs at $\omega=20$.

\FloatBarrier

\section{Frequency-tuned continuous TETs}
\label{sec:scaling}

Within the time interval considered, $\deltalambda(t)$ has two zero crossings at $t_1$ and $t_2$ for $\omega<\omega_{c,1}$ or $\omega>\omega_{c,2}$. They correspond to two linear-crossing TETs and enclose a positive local maximum at $t_p$. As $\omega$ approaches either critical frequency from the side with two crossings, both the crossing-time separation $W=t_2-t_1$ and $\deltalambda(t_p)$ vanish continuously. At the critical frequency, the two crossings coalesce at the critical time $t_c$, where $t_p=t_c$ and $\deltalambda(t_c)=0$. We locate the crossing times by linear interpolation between adjacent samples with opposite signs of $\deltalambda$. The condition $\deltalambda(t_p)=0$ gives $\omega_{c,1}\simeq15.05$ and $\omega_{c,2}\simeq19.61$ for the principal $L=40$ data.

\begin{figure}[!htbp]
\centering
\includegraphics[width=0.88\textwidth]{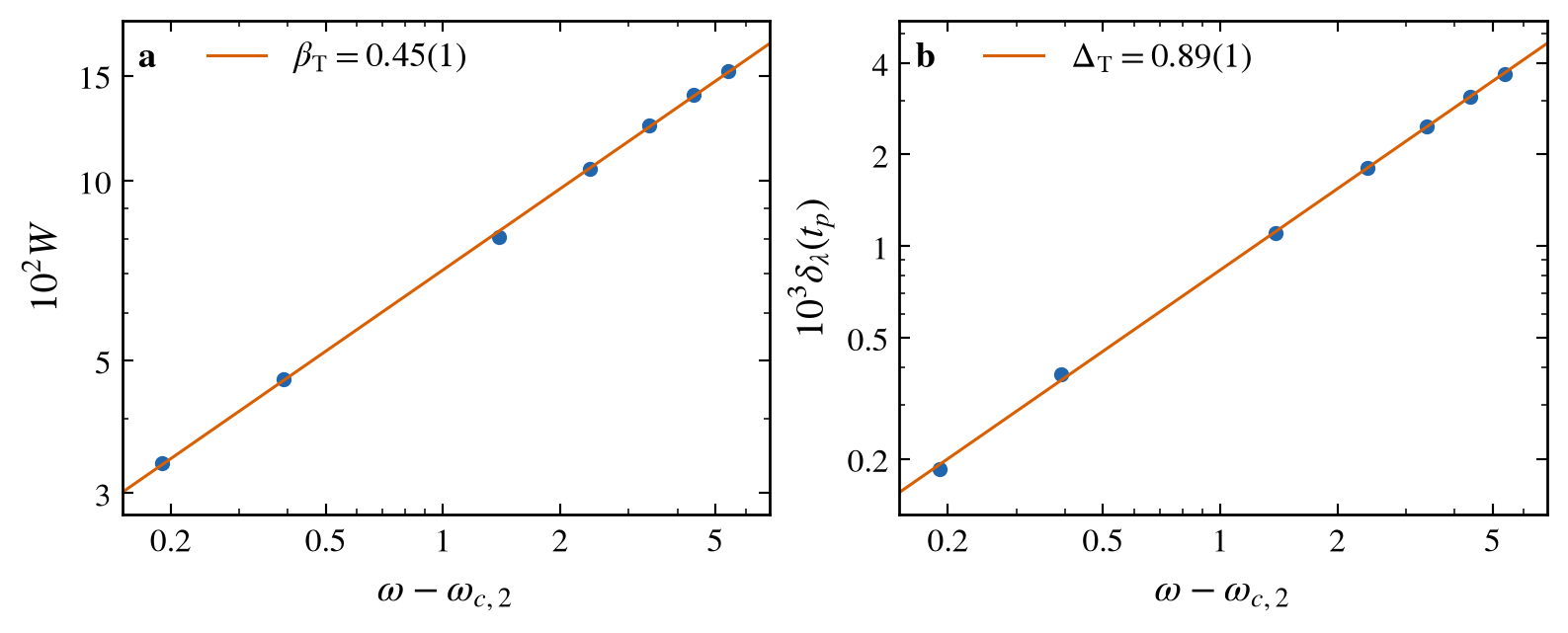}
\caption{Power-law scaling for $\omega>\omega_{c,2}$ in the principal $L=40$ data. (a) Crossing-time separation $W$ and (b) the intervening maximum $\deltalambda(t_p)$ as functions of $\omega-\omega_{c,2}$, with $\omega_{c,2}=19.61$. Symbols show the numerical data, and lines show independent power-law fits.}
\label{fig:sm_upper_scaling}
\end{figure}

Near the upper critical frequency, we use the detuning $r=\omega-\omega_{c,2}>0$. Power-law fits of $W\propto r^{\beta_{\mathrm T}}$ and $\deltalambda(t_p)\propto r^{\Delta_{\mathrm T}}$ give $\beta_{\mathrm T}=0.45(1)$ and $\Delta_{\mathrm T}=0.89(1)$, as shown in Fig.~\ref{fig:sm_upper_scaling}. These independently determined exponents are used in the data collapse in Fig.~3(d) of the main text.

\FloatBarrier
The scaling form introduced in the main text,
\begin{equation}
\deltalambda(t,r)=r^{\Delta_{\mathrm T}}F\!\left(\frac{t-t_p(r)}{r^{\beta_{\mathrm T}}}\right).
\label{eq:sm_scaling}
\end{equation}
contains a time scale $r^{\beta_{\mathrm T}}$ and a scale $r^{\Delta_{\mathrm T}}$ for $\deltalambda$. At the critical frequency, the local profile obeys $\deltalambda(t,0)\propto|t-t_c|^{\delta_{\mathrm T}}$. Rescaling time by $r^{\beta_{\mathrm T}}$ therefore rescales $\deltalambda$ by $r^{\beta_{\mathrm T}\delta_{\mathrm T}}$. Comparing this with the prefactor $r^{\Delta_{\mathrm T}}$ in Eq.~\eqref{eq:sm_scaling} gives
\begin{equation}
\Delta_{\mathrm T}=\beta_{\mathrm T}\delta_{\mathrm T}.
\label{eq:sm_exponent_relation}
\end{equation}
The exponents at $\omega_{c,2}$, together with $\delta_{\mathrm T}=1.97(1)$, satisfy this relation within their uncertainties. The scaling of the crossing-time separation also gives $dW/d\omega\propto r^{\beta_{\mathrm T}-1}$, which diverges as $r\to0^+$.

\begin{figure}[!htbp]
  \centering
  \includegraphics[width=0.98\textwidth]{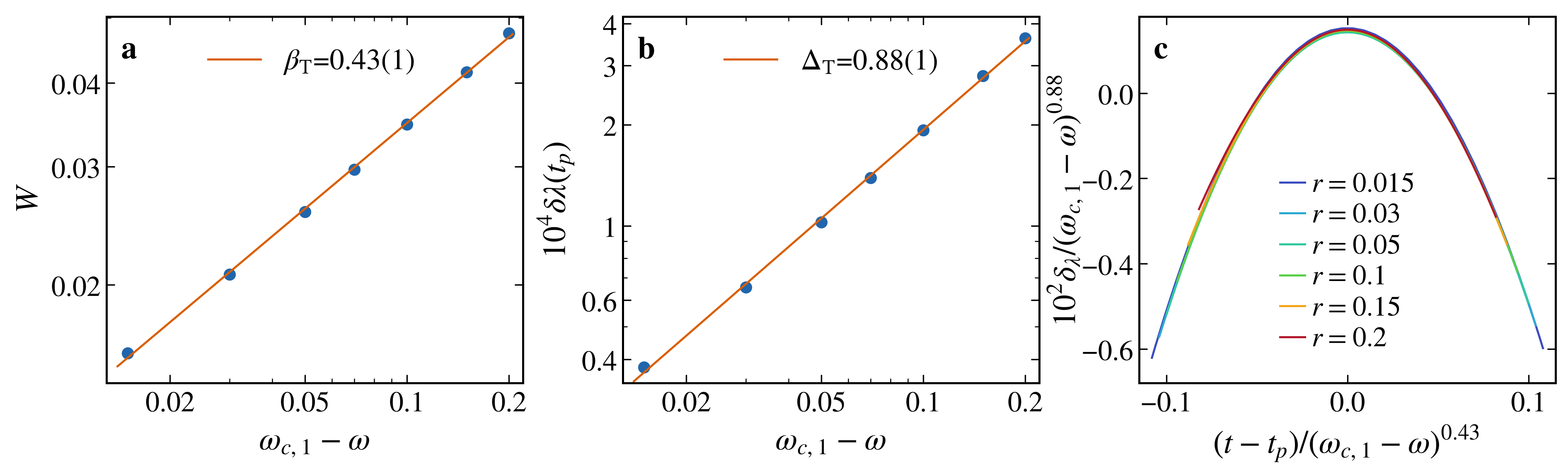}
\caption{Lower-endpoint scaling for $L=40$, $L_A=10$, $J_2=-0.5$, $h_0=4$, and $\hini=7$, using $\Delta t=0.001$ and $\omega_{c,1}=15.05$. (a) Crossing-time separation $W$ and (b) peak height $\deltalambda(t_p)$ versus $r=\omega_{c,1}-\omega$ for frequencies in $ r\le0.20$. Lines are independent power-law fits giving $\beta_{\mathrm T}=0.43(1)$ and $\Delta_{\mathrm T}=0.88(1)$. (c) Collapse of representative temporal profiles using these fitted powers.}
  \label{fig:sm_lower_scaling}
\end{figure}

At the lower critical frequency, we use $r=\omega_{c,1}-\omega>0$ with $\omega_{c,1}=15.05$. Independent fits to the frequencies in $r\le0.20$ give $\beta_{\mathrm T}=0.43(1)$ from $W$ and $\Delta_{\mathrm T}=0.88(1)$ from $\deltalambda(t_p)$. These fitted powers produce the profile collapse shown in Fig.~\ref{fig:sm_lower_scaling}. Their ratio, $\Delta_{\mathrm T}/\beta_{\mathrm T}=2.05(5)$, is also close to the local touching exponent $\delta_{\mathrm T}=1.97(1)$.

\FloatBarrier
\section{Dependence on model parameters and system sizes}
\label{sec:robustness}

\begin{figure}[!htbp]
\centering
\includegraphics[width=0.47\textwidth]{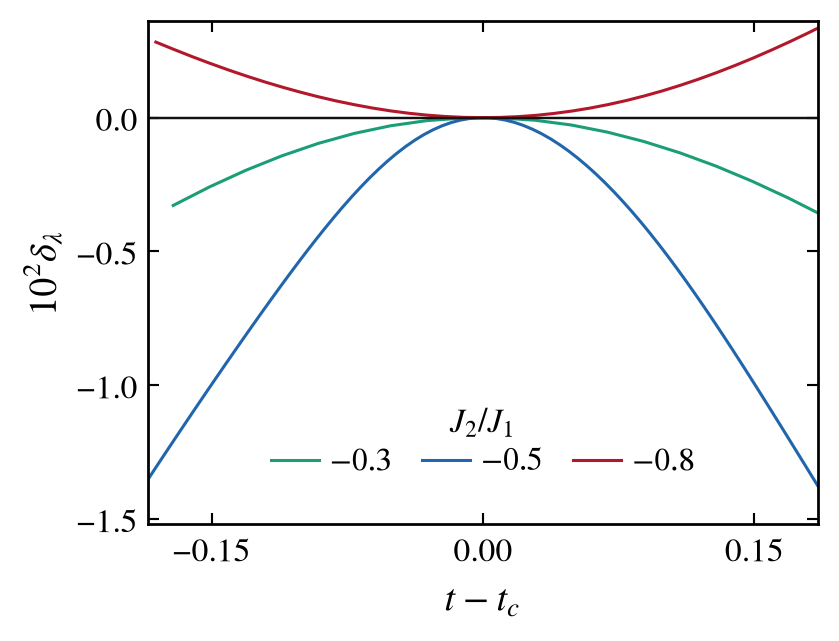}
\caption{Interpolated $\deltalambda(t)$ profiles at critical touchings for $(J_2/J_1,h_0/J_1)=(-0.3,2)$, $(-0.5,4)$, and $(-0.8,2)$, with critical frequencies $\omega_c\simeq5.40$, $15.05$, and $6.29$, respectively. Time is measured relative to the corresponding critical time $t_c$. All calculations use $L=40$, $L_A=10$, and parity-preserving initial states.}
\label{fig:sm_interactions}
\end{figure}

Figure~\ref{fig:sm_interactions} compares $\deltalambda(t)$ at critical touchings for three sets of interaction and drive parameters. Each curve is obtained by interpolating between frequencies that bracket the corresponding critical frequency. Although the critical frequency and the time dependence of $\deltalambda$ vary with the parameters, both the tangential touching and its critical scaling persist in every case examined.

\FloatBarrier
For the principal parameters $J_2=-0.5$, $h_0=4$, and $\hini=7$, we also varied the system and subsystem sizes. At fixed $L_A=10$, the lower and upper critical frequencies lie within $15.00$--$15.06$ and $19.6$--$19.8$, respectively, for $L=20$, $30$, and $40$. At fixed $L=40$, the same intervals apply to $L_A=8$; for $L_A=6$, they shift slightly to $15.06$--$15.10$ and $19.4$--$19.6$. Thus, the two critical touchings and the intervening frequency interval in which no TET occurs are stable for all sizes examined.

\FloatBarrier
\section{Subsystem-parity sector weights}

\begin{figure}[!htbp]
  \centering
  \includegraphics[width=0.78\textwidth]{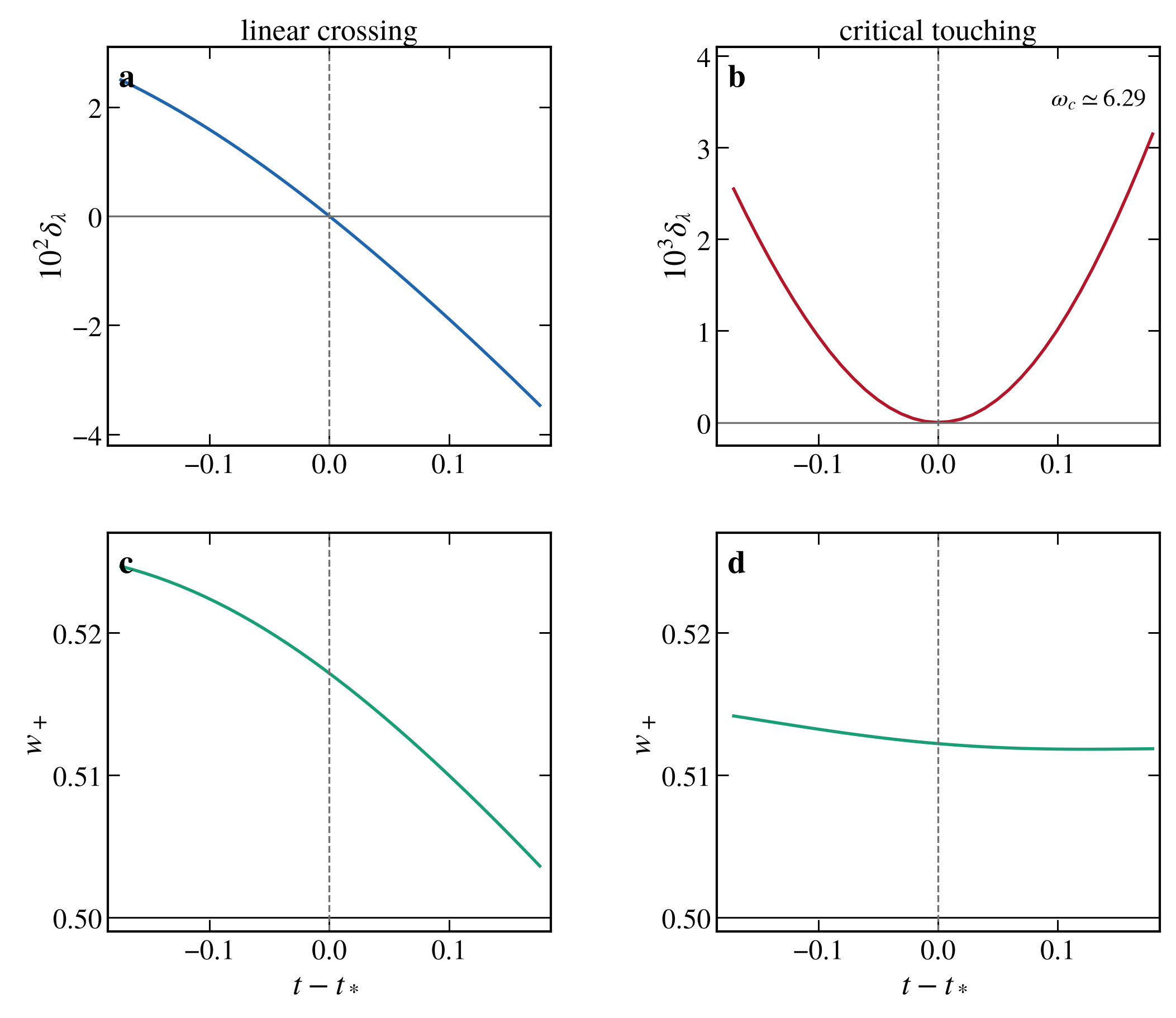}
\caption{Comparison of $\deltalambda$ and $w_+$ for $L=40$, $L_A=10$, $J_2=-0.8$, and $h_0=2$. Panels (a,c) show a linear-crossing TET at $\omega=5.5$. Panels (b,d) show a critical touching reconstructed by interpolation at $\omega_c\simeq6.29$. Time is measured relative to the crossing time in (a,c) and the critical time $t_c$ in (b,d).}
  \label{fig:sm_wplus_comparison}
\end{figure}

For a state with definite global Ising parity, $[\rho_A,\mathcal P_A]=0$, and $\rho_A$ is block diagonal in the subsystem-parity sectors. The total weight of the even-parity block is
\begin{equation}
 w_+(t)=\mathrm{Tr}\!\left[\Pi_+\rho_A(t)\right]
 =\frac{1+\langle\mathcal P_A\rangle}{2},
 \qquad \Pi_+=\frac{1+\mathcal P_A}{2}.
\end{equation}
The quantity $\deltalambda$ compares only the largest Schmidt value in each parity sector, whereas $w_+$ is the sum of all Schmidt values in the even sector. Consequently, the condition $\lambda_+=\lambda_-$ does not imply $w_+=1/2$. Figure~\ref{fig:sm_wplus_comparison} illustrates this distinction. At the linear-crossing TET, $\deltalambda$ changes sign while $w_+$ approaches to $0.5$ and equals $0.517$ at the crossing. At the critical touching, $\deltalambda$ reaches zero without changing sign, while $w_+$ remains smooth and equals $0.512$ at $t_c$.

\FloatBarrier
\section{Floquet effective Hamiltonian}

Write $H(t)=H_I+V\cos(\omega t)$ with $V=-h_0\sum_iS_i^x$, and define the stroboscopic Floquet Hamiltonian by $U(T,0)=e^{-iH_FT}$, where $T=2\pi/\omega$. We choose $t=0$ at the maximum of the cosine drive. For this time-symmetric choice of stroboscopic origin, $H(T-t)=H(t)$, so the $O(\omega^{-1})$ correction to $H_F$ vanishes. The Floquet--Magnus expansion through $O(\omega^{-2})$ gives~\cite{Eckardt2015}
\begin{equation}
 H_F^{(2)}=H_I+\frac{1}{\omega^2}
 \left\{-\left[H_I,[H_I,V]\right]
 +\frac14\left[V,[H_I,V]\right]\right\}.
 \label{eq:sm_hf_commutator}
\end{equation}

For the open chain, define the local longitudinal-field operator
\begin{equation}
 B_i=\sum_{\substack{r=1,2;\ s=\pm1\\1\leq i+sr\leq L}}
 J_rS_{i+sr}^z.
\end{equation}
The spin commutation relations give
\begin{align}
 -[H_I,[H_I,V]]
 & =h_0\sum_iS_i^xB_i^2,
 \\
 \frac14[V,[H_I,V]]
 & =\frac{h_0^2}{2}\sum_{r=1,2}J_r
 \sum_{i=1}^{L-r}
 \left(S_i^zS_{i+r}^z-S_i^yS_{i+r}^y\right).
\end{align}
These identities reproduce the effective Hamiltonian in the main text. The $S_i^zS_{i+r}^z$ contribution renormalizes the Ising couplings, while the $S_i^yS_{i+r}^y$ contribution generates $S^yS^y$ interactions. Since $(S_j^z)^2=1/4$, the $S_i^xB_i^2$ term contains an effective transverse-field term and three-spin interactions $S_j^zS_i^xS_k^z$. All generated terms preserve global Ising parity, while the $S^yS^y$, transverse-field, and three-spin terms do not commute with $H_I$.

\begin{figure}[!htbp]
  \centering
  \includegraphics[width=0.90\textwidth]{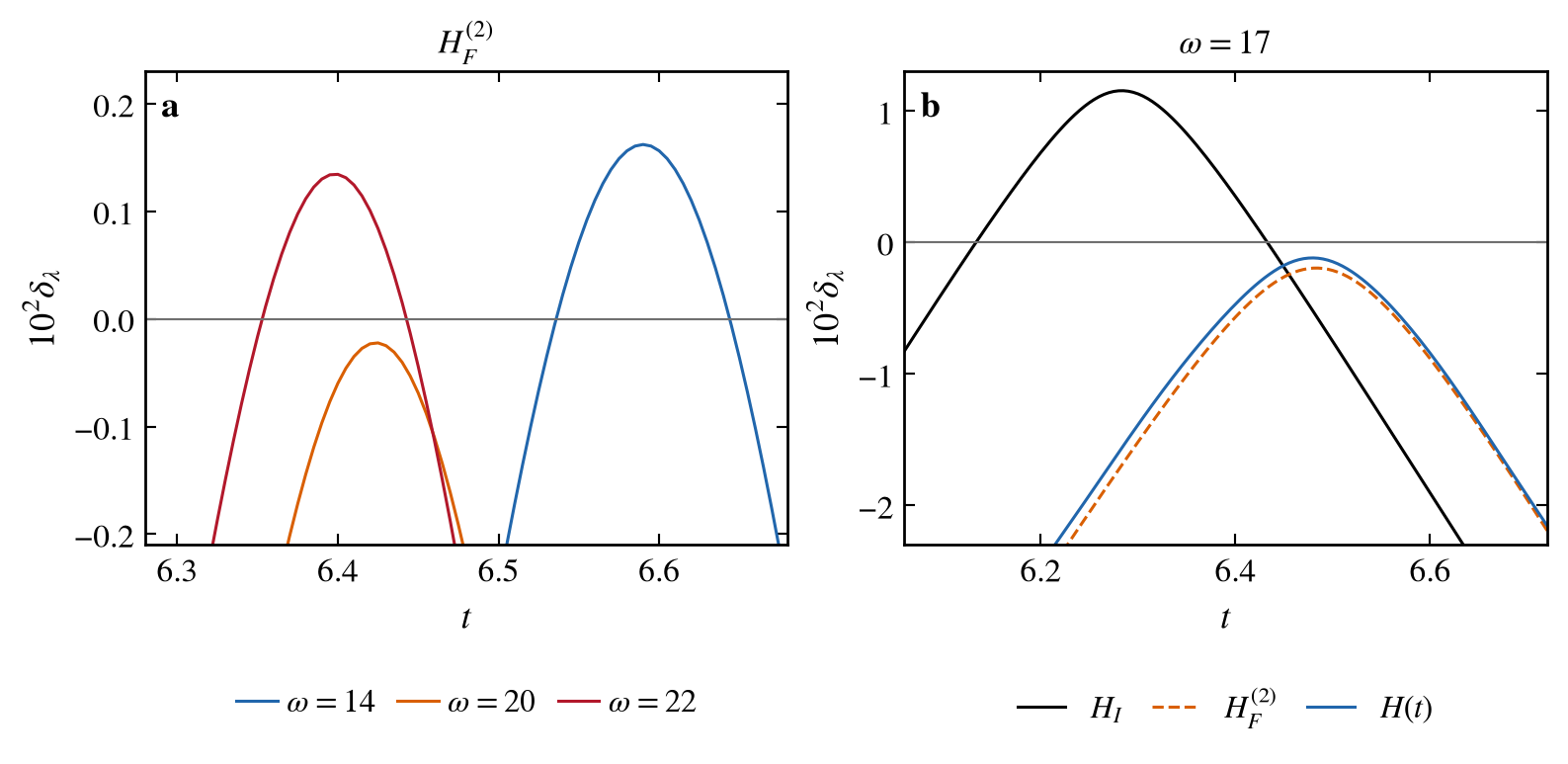}
\caption{Comparison with the second-order Floquet Hamiltonian for $L=40$, $L_A=10$, $J_2=-0.5$, $h_0=4$, and the $\hini=7$ initial ground state. (a) $\deltalambda(t)$ under $H_F^{(2)}$ below the lower critical frequency, between the two critical frequencies, and above the upper critical frequency. (b) Comparison of $H_I$, $H_F^{(2)}$, and the full driven evolution at $\omega=17$.}
  \label{fig:sm_floquet}
\end{figure}

Figure~\ref{fig:sm_floquet}(a) shows that $H_F^{(2)}$ reproduces the disappearance and reappearance of the two zero crossings, with critical frequencies $\omega_{c,1}^{(F)}\simeq14.71$ and $\omega_{c,2}^{(F)}\simeq20.28$. At $\omega=17$, $H_I$ produces two zero crossings of $\deltalambda(t)$, whereas the Schmidt gap $|\deltalambda(t)|$ remains nonzero under both $H_F^{(2)}$ and the full driven evolution in the displayed time interval [Fig.~\ref{fig:sm_floquet}(b)]. The $O(\omega^{-2})$ terms therefore reproduce the finite frequency interval in which no TET occurs. The difference between the critical frequencies of $H_F^{(2)}$ and the full driven evolution can arise from higher-order Floquet terms and micromotion omitted here~\cite{Goldman2014,Eckardt2015}.

\FloatBarrier

\end{document}